\documentclass{PoS}
\usepackage{latexsym,amssymb,amsfonts,amsmath}

\newcommand{\f}[2]{\frac{#1}{#2}}

\newcommand{\la}{\langle}
\newcommand{\ra}{\rangle}

\title{Critical behaviour in the QCD Anderson transition}

\ShortTitle{Critical behaviour in the QCD Anderson transition }

\author{\speaker{Matteo Giordano}\thanks{Supported by the Hungarian
    Academy of Sciences under ``Lend\"ulet'' grant
    No. LP2011-011. TGK and FP acknowledge partial support by the EU
    Grant (FP7/2007 -2013)/ERC No. 208740. We also thank the
    Budapest-Wuppertal group for allowing us to use their code to
    generate the gauge configurations.}\\
  Institute for Nuclear Research of the Hungarian Academy of
  Sciences,\\
  Bem t\'er 18/c, H-4026 Debrecen, Hungary\\
  E-mail: \email{giordano@atomki.mta.hu}}

\author{Tam\'as G.\ Kov\'acs\footnotemark[2]\\
  Institute for Nuclear Research of the Hungarian Academy of
  Sciences,\\
  Bem t\'er 18/c, H-4026 Debrecen, Hungary\\
  E-mail: \email{kgt@atomki.mta.hu}}
\author{Ferenc Pittler\footnotemark[2]\\
  Institute for Nuclear Research of the Hungarian Academy of
  Sciences,\\
  Bem t\'er 18/c, H-4026 Debrecen, Hungary\\
  E-mail: \email{pittler@atomki.mta.hu}}

\abstract{We study the Anderson-type localisation-delocalisation
  transition found previously in the QCD Dirac spectrum at high
  temperature. Using high statistics QCD simulations with $N_f=2+1$
  flavours of staggered quarks, we discuss how the change in the
  spectral statistics depends on the volume, the temperature and the
  lattice spacing, and we speculate on the possible universality of
  the transition from Poisson to Wigner-Dyson in the spectral
  statistics. Moreover, we show that the transition is a genuine phase
  transition: at the mobility edge, separating localised and
  delocalised modes, quantities characterising the spectral statistics
  become non-analytic in the thermodynamic limit. Using finite size
  scaling we also determine the critical exponent of the correlation
  length, and we speculate on possible extensions of the universality
  of Anderson transitions.}

\FullConference{31st International Symposium on Lattice Field Theory -
  LATTICE 2013\\ 
  July 29 - August 3, 2013\\
  Mainz, Germany}

\begin{document}

\section{Introduction}

\begin{figure}[t]
  \centering
  \includegraphics[width=0.49\textwidth]{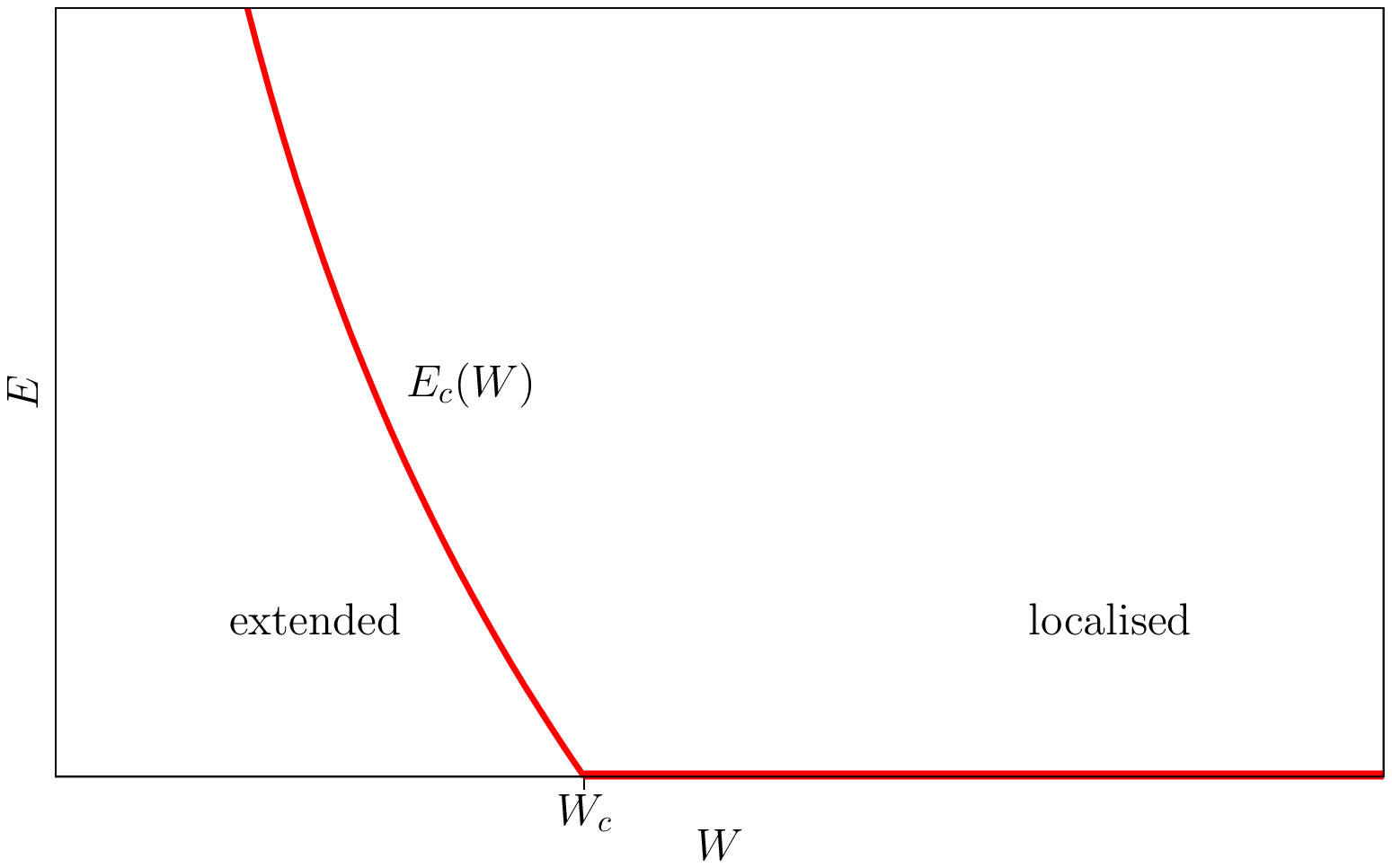}\hspace{\stretch{1}}\includegraphics[width=0.49\textwidth]{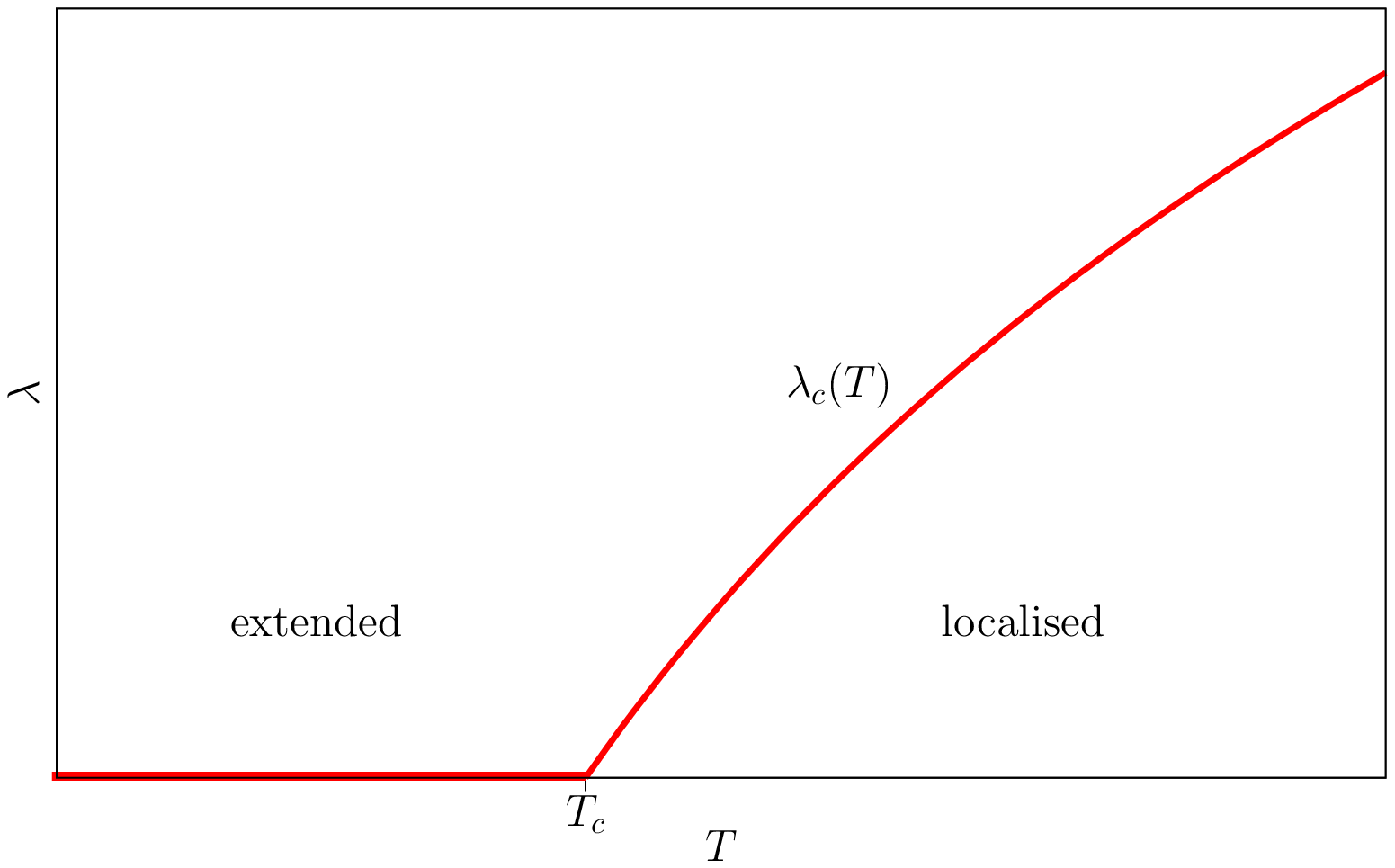} 
  \caption{Schematic phase diagram of the 3D Anderson model (left) and QCD
    (right) in the disorder/eigenvalue plane.}
  \label{fig:1}
\end{figure}

It is well known that the properties of the low-lying modes of the
Dirac operator are intimately related to the behaviour of QCD under
chiral symmetry transformations, as clearly exemplified by the 
Banks-Casher relation~\cite{BC}.
In particular, it has been realised in recent years that their
localisation properties 
change completely across the
chiral transition/crossover. 
While below the critical temperature 
$T_c$ all the eigenmodes are delocalised, it has been
shown~\cite{GGO,KGT,KP,KP2} that above $T_c$ the low-lying ones, up to
some critical point $\lambda_c$, become localised; modes above
$\lambda_c$ remain delocalised. Initially the evidence for this was
mainly obtained in the {\it quenched} approximation and/or for the
$SU(2)$ gauge group, but recently this scenario has been demonstrated
in full QCD~\cite{KP2}, by studying the spectrum of the staggered
Dirac operator in numerical simulations of lattice QCD with $N_f=2+1$
flavours of quarks at physical masses~\cite{BW}. An improved study,
with much higher statistics and larger lattice volumes, has been
presented at this conference~\cite{feri}. 

The presence of a transition from localised to delocalised modes in
the spectrum, as the one found in QCD above $T_c$, is a well known
phenomenon in condensed matter physics, and it represents the main
feature of the celebrated Anderson model~\cite{Anderson} in three
dimensions. The Anderson model aims at a description of electrons in a
``dirty'' conductor, by mimicking the effect of impurities through
random interactions. In its lattice version, the model is obtained by
adding a random on-site potential to the usual tight-binding
Hamiltonian, 
\begin{equation}
  \label{eq:andersonmodel}
  H = \sum_n \varepsilon_n| n \ra \la n | + \sum_{n}\sum_{\mu=1}^3 |n +
  \hat\mu\ra\la n | + |n \ra\la n + \hat\mu|\,,
\end{equation}
where $| n \ra$ denotes a state localised on the lattice site $n$, and
$\varepsilon_n$ are random variables drawn from some distribution, whose 
width $W$ measures the amount of disorder, i.e., of impurities in the
system. The phase diagram of this model is sketched in
Fig.~\ref{fig:1}. While for $W=0$ all the eigenmodes are delocalised,
localised modes appear at the band edge as soon as the random
interaction is switched on. The critical energy $E_c$ separating
localised and delocalised modes is called ``mobility  edge'', and its
value depends on the amount of disorder, $E_c=E_c(W)$. As $W$
increases, $E_c$ moves towards the center of the band, and above a
critical disorder $W_c$ all the modes become localised. From the
physical point of view, this signals a transition of the system from
metal to insulator. 

In Fig.~\ref{fig:1} we also sketch a schematic phase diagram for
QCD. Here the role of disorder is played by the temperature, while the
energy is replaced by the eigenvalue of the Dirac operator. Localised
modes are present in the low end of the spectrum above $T_c$, up to
the ``mobility edge'' $\lambda_c(T)$. Around the critical temperature 
$\lambda_c$ vanishes~\cite{KP2}, and below $T_c$ all the modes are
extended.  

In both models, localised modes appear where the spectral density
is small. One then expects that they are not easily mixed
by the fluctuations of the random interaction, which in turn suggests
that the corresponding eigenvalues are statistically independent,
obeying Poisson statistics. On the other hand, eigenmodes remain
extended in the region of large spectral density also in the presence
of disorder, and so one expects them to be basically freely mixed by
fluctuations. The corresponding eigenvalues are then expected
to obey the Wigner-Dyson statistics of Random Matrix Theory (RMT).  
This connection between localisation of eigenmodes and eigenvalue
statistics provides a convenient way to detect the
localisation/delocalisation transition and study its critical
properties.  

The transition from Poisson to RMT behaviour in the local spectral 
statistics is most simply studied by means of the so-called unfolded level
spacing distribution (ULSD). Unfolding consists essentially in a local
rescaling of the eigenvalues to have unit spectral density throughout the 
spectrum. The ULSD gives the probability distribution of the
difference between two consecutive eigenvalues of the Dirac operator 
normalised by the local average level spacing. The ULSD is known
analytically for both kinds of behaviour: in the case of Poisson
statistics it is a simple exponential, while in the case of
RMT statistics it is very precisely approximated by the so-called ``Wigner
surmise'' for the appropriate symmetry class, which for QCD is the
unitary class,  
\begin{equation}
  \label{eq:ulsd}
 P_{\rm Poisson}(s)=e^{-s} \,, \qquad P_{\rm RMT}(s)=\f{32}{\pi^2} s^2
 e^{-\f{4}{\pi}s^2}  \,. 
\end{equation}
Rather than using the full distribution to characterise the local spectral
statistics, it is more practical to consider a single parameter of the
ULSD. Any such quantity, having different values for Poisson and RMT
statistics, can be used to detect the Poisson/RMT transition. In our
study, we used the integrated ULSD and the second moment of the
ULSD, 
\begin{equation}
  \label{eq:il_s2}
   I_\lambda = \int_0^{s_0} ds\,
P_\lambda(s)\,,\quad s_0\simeq 0.508\,, 
\qquad \la s^2 \ra_\lambda= \int_0^\infty ds\, P_\lambda(s)\, s^2 \,,
\end{equation}
defined locally in the spectrum. The choice of $s_0$ was made in order
to maximise the difference between the Poisson and RMT predictions,
namely $I_{\rm Poisson}\simeq 0.398$ and $I_{\rm RMT}\simeq 0.117$;
as for the second moment, the predictions are $\la s^2 \ra_{\rm
  Poisson}=2$ and $\la s^2 \ra_{\rm RMT}=3\pi/8$.

\section{Numerical results}

\begin{figure}[t]
  \centering
  \includegraphics[width=0.49\textwidth]{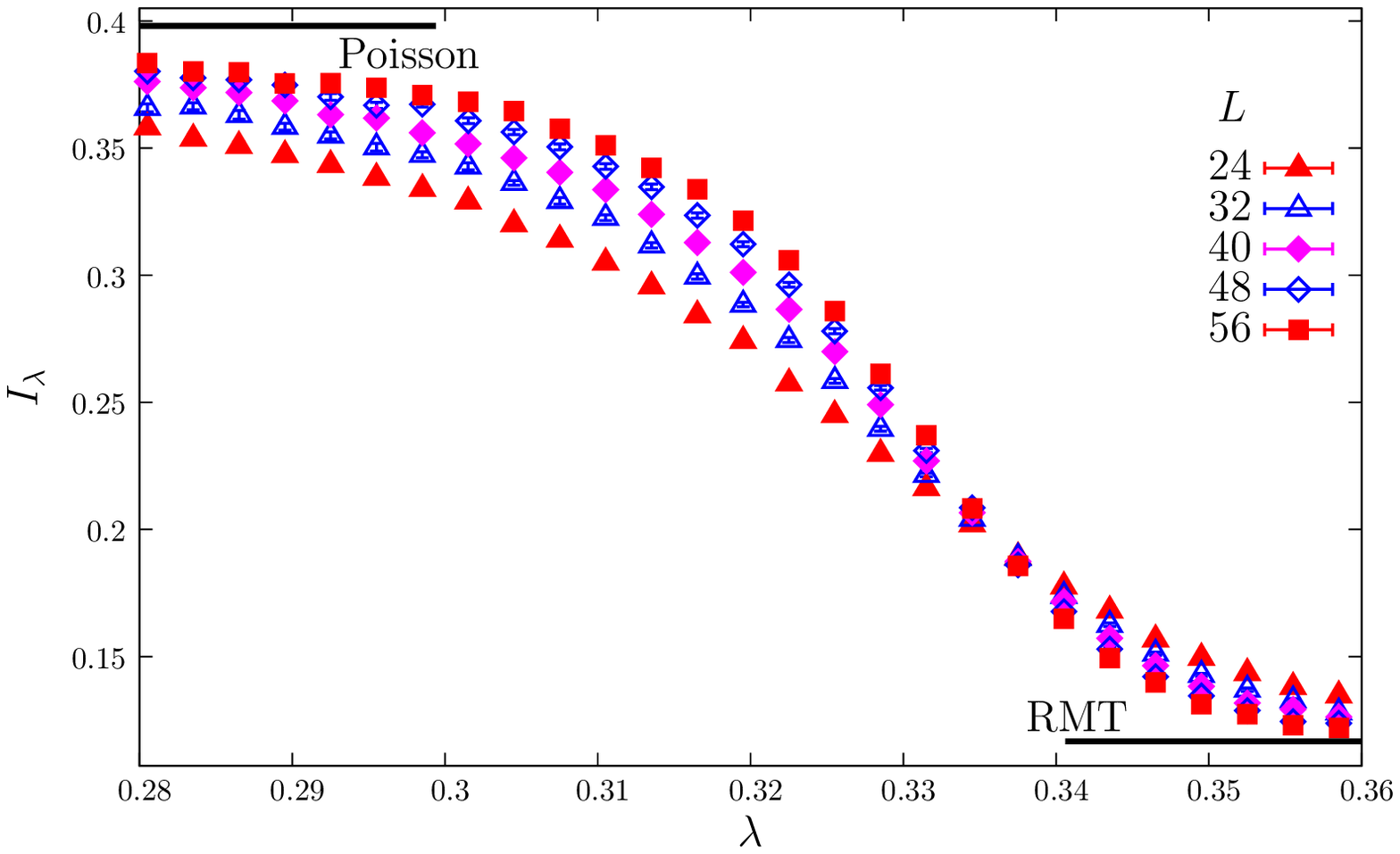}\hspace{\stretch{1}}\includegraphics[width=0.49\textwidth]{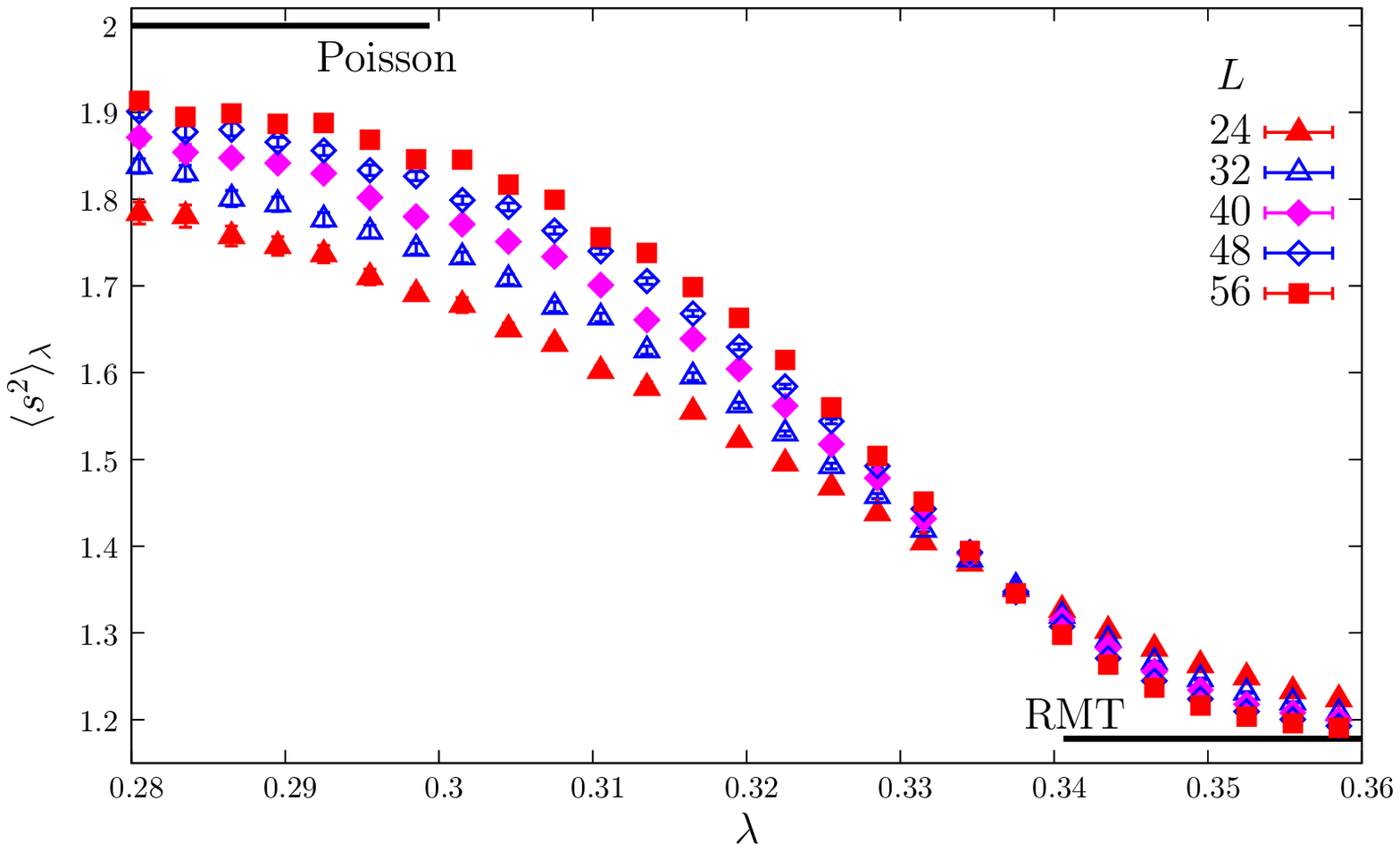}
  \caption{Integrated ULSD (left) and second moment of the ULSD
    (right), computed locally along the spectrum, for several lattice
    sizes. Here $\Delta\lambda=3\cdot 10^{-3}$.} 
  \label{fig:2}
\end{figure}

The results presented here are based on simulations of lattice QCD using
a Symanzik-improved gauge action and $2+1$ flavours of stout smeared
staggered fermions, with quark masses at physical values~\cite{BW}. We used
a lattice of fixed temporal extension $N_t=4$ at $\beta=3.75$, corresponding to
lattice spacing $a=0.125~{\rm fm}$ and physical temperature $T=394~{\rm
  MeV}\simeq 2.6 T_c$. For different choices of spatial size
$L=24,28,32,36,40,44,48,56$ in lattice units, we collected large
statistics for eigenvalues and eigenvectors of the staggered Dirac
operator in the relevant spectral range - see Ref.~\cite{feri} for
more details. Here and in the following the eigenvalues $\lambda$ are
expressed in lattice units. Unfolding was done by ordering all the
eigenvalues obtained on all the configurations (for a given volume)
according to their magnitude, and replacing them by their rank order
divided by the total number of configurations. We then computed
locally the integrated ULSD and the second moment of the ULSD, by
dividing the spectrum in small bins of size $\Delta\lambda$, computing
the observables in each bin, and assigning the resulting value to the
average value of $\lambda$ in each bin. We used several values for
$\Delta\lambda$, ranging from $1\cdot 10^{-3}$ to $6\cdot 10^{-3}$. 

In Fig.~\ref{fig:2} we show the integrated ULSD $I_\lambda$ and the second
moment of the ULSD $\la s^2 \ra_\lambda$, for several values of the
spatial volume. A transition from Poisson to RMT is clearly visible,
and moreover it gets sharper and sharper as the volume of the lattice
is increased. This suggests that the transition becomes a true phase
transition in the thermodynamic limit.

\section{Finite size scaling}

To check if the Poisson/RMT transition in the spectral statistics
(i.e., the localisation/de\-lo\-ca\-li\-sation transition) is a genuine,
Anderson-type phase transition, we have performed a finite size 
scaling analysis, along the lines of Refs.~\cite{HS,SSSLS,SP}.
The Anderson transition is a second-order phase transition, with the
characteristic length of the system $\xi_\infty$ diverging at the 
critical point $\lambda_c$ like $\xi_\infty(\lambda)\sim 
|\lambda-\lambda_c|^{-\nu}$. To determine the critical exponent $\nu$ and
the critical point $\lambda_c$, one picks a dimensionless quantity
$Q(\lambda,L)$, measuring some local statistical properties of the
spectrum, and having different thermodynamic limits on the two sides
of the transition (and possibly at the critical point), i.e., 
\begin{equation}
  \label{eq:observable}
      \lim_{L\to\infty} Q(\lambda,L) = \left\{
          \begin{aligned}
            &Q_{\rm Poisson} &&& &\hspace{-0.40cm} \lambda<\lambda_c &&&
            &\hspace{-0.40cm}\text{(localised)}\,,\\  
            &Q_c &&& &\hspace{-0.40cm} \lambda=\lambda_c &&&
            &\hspace{-0.40cm}\text{(critical)}\,,\\ 
            &Q_{\rm RMT} &&& &\hspace{-0.40cm} \lambda>\lambda_c &&&
            &\hspace{-0.40cm}\text{(delocalised)}\,.
          \end{aligned}
        \right.
\end{equation}
As the notation suggests, $Q(\lambda,L)$ is computed on a lattice of 
linear size $L$. For large enough volume, and close to the
critical point, finite size scaling suggests that the
dependence of $Q$ on $L$ is of the form $Q(\lambda,L) =
f(L/\xi_\infty(\lambda))$. As $Q(\lambda,L)$ is
analytic in $\lambda$ for any finite $L$, we must have
\begin{equation}
  \label{eq:fss}
Q(\lambda,L)=F(L^{1/\nu}(\lambda-\lambda_c))  \,,
\end{equation}
with $F$ analytic. Here we have assumed that corrections to one-parameter
scaling can be neglected.


\begin{figure}[t]
  \centering
\includegraphics[width=0.31\textwidth]{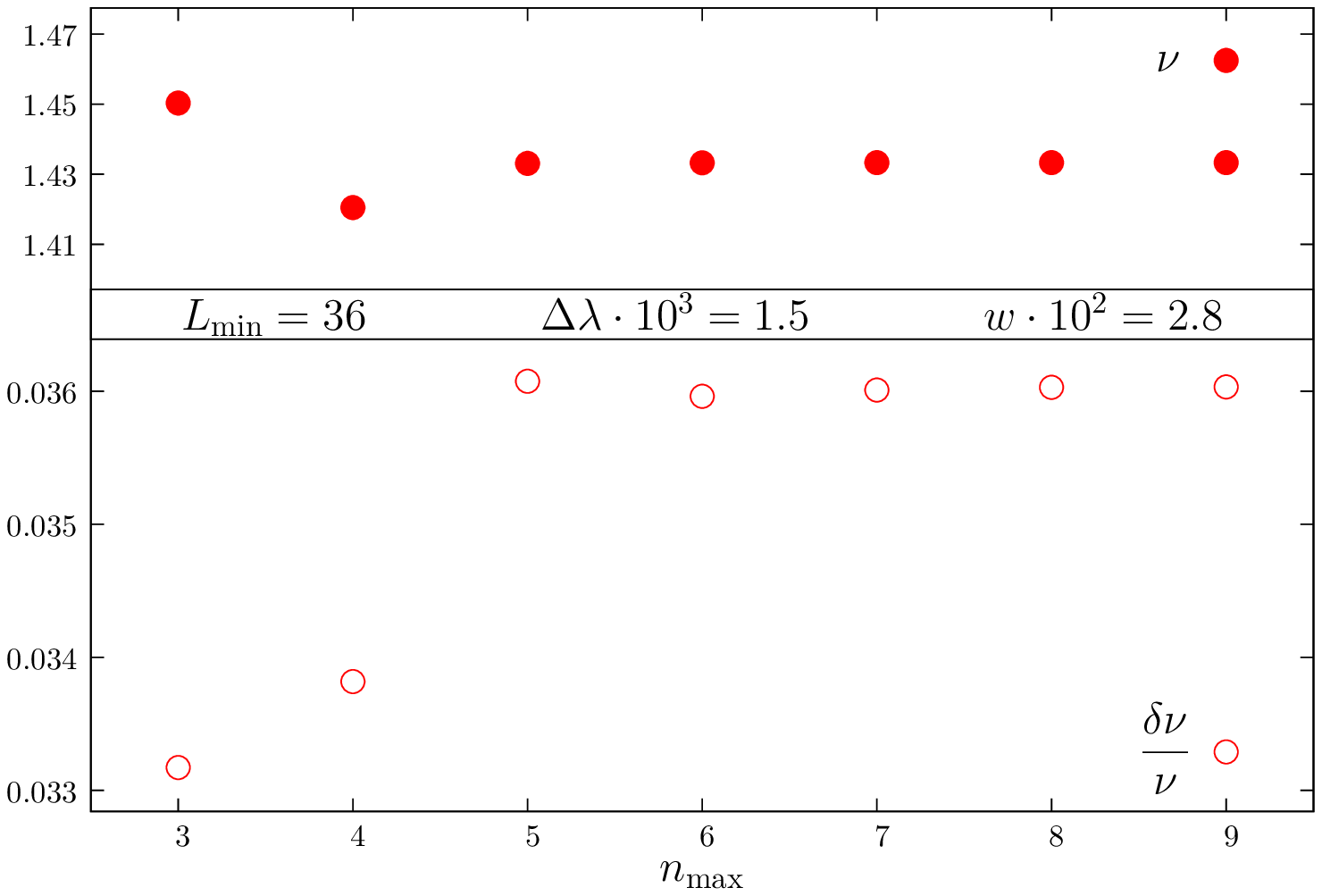}\hspace{\stretch{3}}\includegraphics[width=0.33\textwidth]{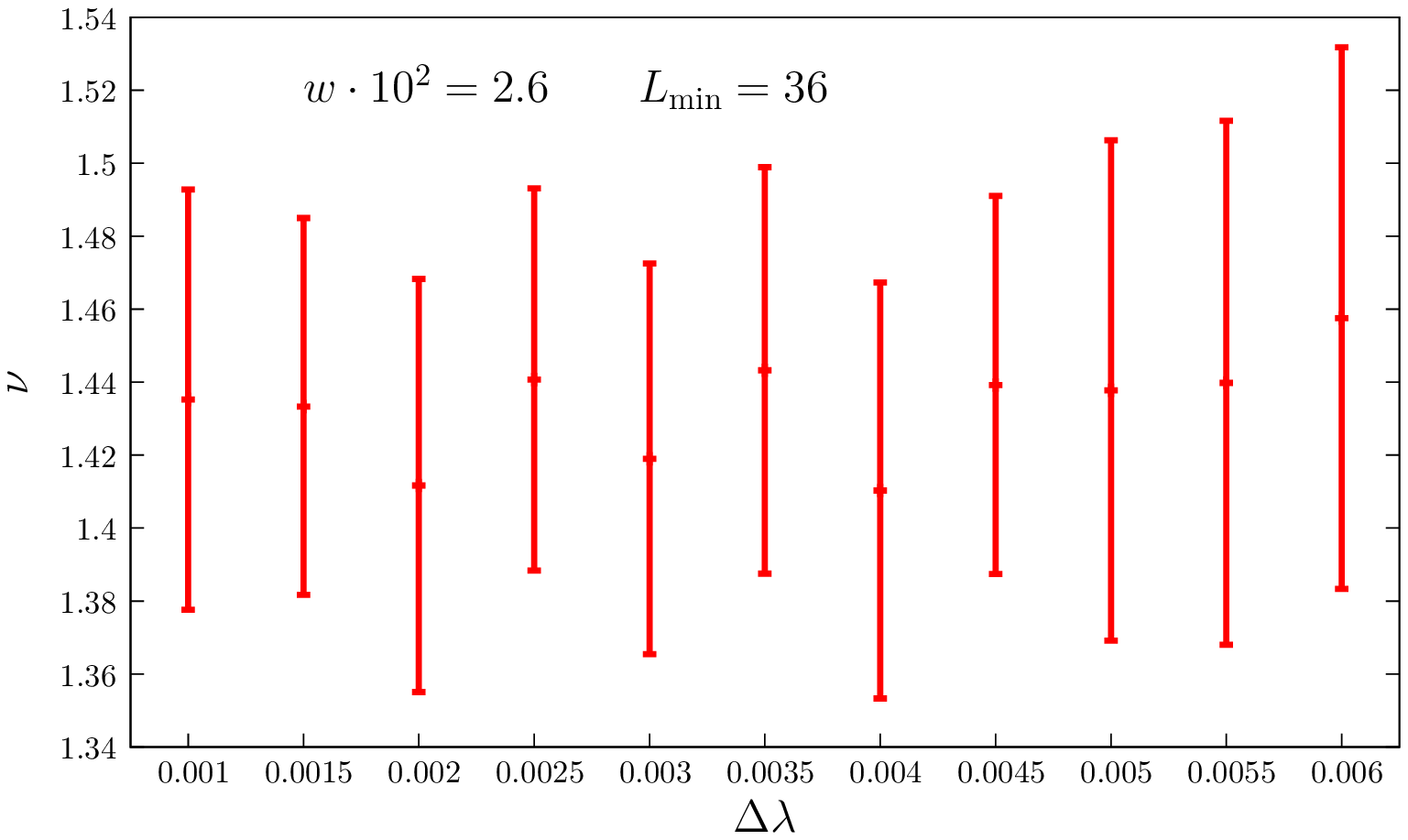}
\hspace{\stretch{1}}  \includegraphics[width=0.33\textwidth]{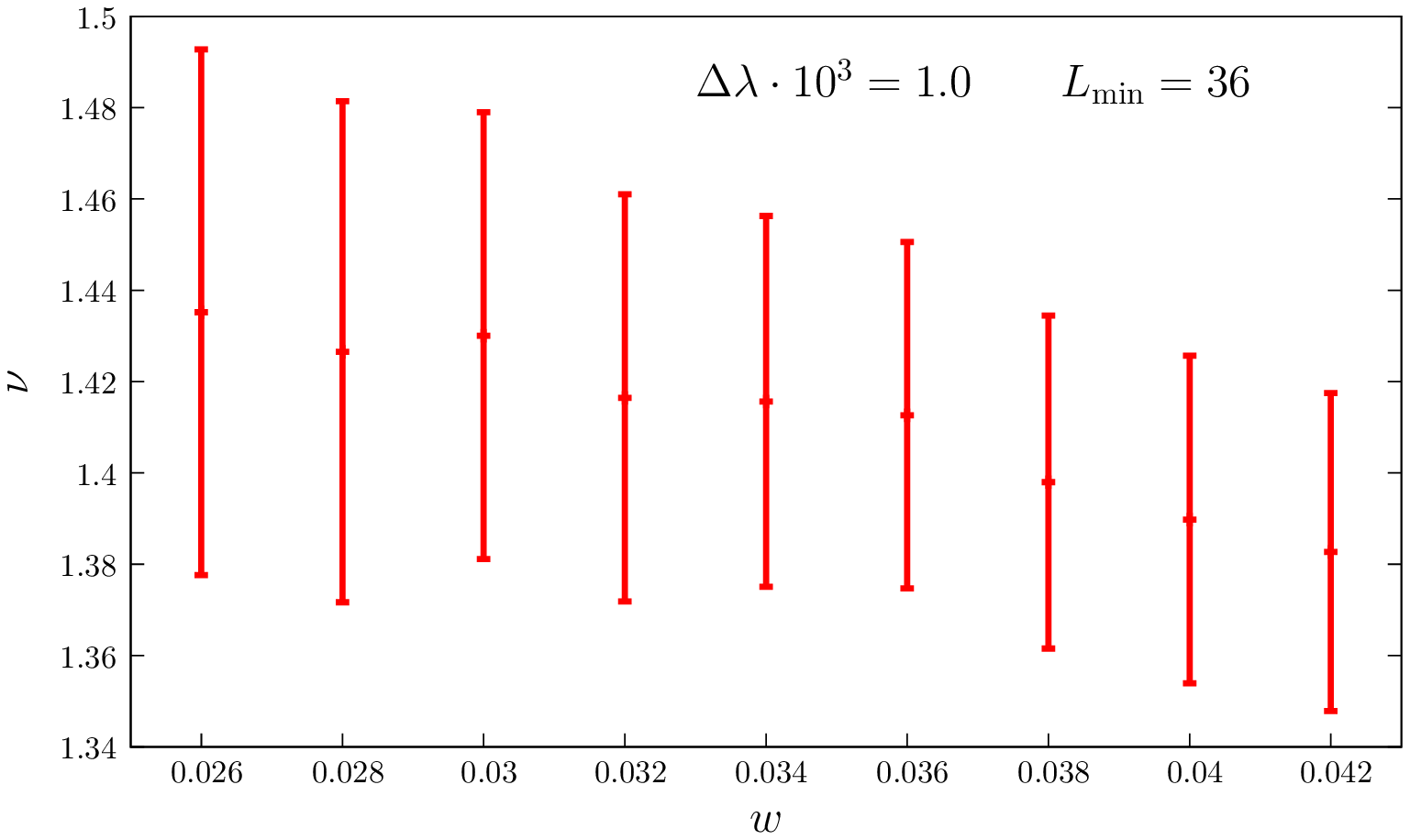}
  \caption{Dependence of the fitted value of $\nu$ and corresponding
    relative error as a function of the number of terms $n_{\rm max}$,
    in the case of $L_{\rm min}=36$, $\Delta\lambda\cdot 10^3=1.5$ and
    $w\cdot 10^2=2.8$ (left).
    Dependence of the fitted value of $\nu$ on the bin size
    $\Delta\lambda$ for the smallest fitting
    range (center) and on the width $w$ of
  the fitting range for the smallest bin size (right). Here
  $L_{\rm min}=36$. } 
  \label{fig:binwidth_dep}
\end{figure}

If one determines $\lambda_c$ and $\nu$ correctly, the numerical data
for $Q(\lambda,L)$ obtained for different lattice sizes should collapse
on a single curve, when plotted against the scaling variable
$L^{1/\nu}(\lambda-\lambda_c)$. We then proceeded as follows:
expanding the scaling function $F$ in powers of $\lambda-\lambda_c$,
one gets 
\begin{equation}
  \label{eq:fss_expansion}
 Q(\lambda,L)=\sum_{n=0}^{\infty} F_{n}\,L^{n/\nu}(\lambda-\lambda_c)^n \,.
\end{equation}
By truncating the series to some $n_{\rm max}$ and performing a
fit to the numerical data, using several volumes at a
time, one can then determine $\nu$ and $\lambda_c$, together with the
first few coefficients $F_n$. 
For our purposes, the best quantity turned out to be the integrated ULSD
$I_\lambda$. Our fitter was based on the MINUIT
library~\cite{JR}. Statistical errors were determined by means of a
jackknife analysis. To check for finite size effects, we repeated the
fit using only data from lattices of size $L\ge L_{\rm min}$ for
increasing $L_{\rm min}$. 

Systematic effects due to the truncation of the series for the scaling
function, Eq.~\eqref{eq:fss_expansion}, are controlled by including more
and more terms in the series, and checking how the results change. In
order to circumvent the numerical 
instability of polynomial fits of large order, we resorted to the
technique of constrained fits~\cite{LCDHMMT}. The basic idea of constrained
fits is to use the available information to constrain the 
values of the fitting parameters. In our case, they are needed only to
avoid that the polynomial coefficients of higher order take unphysical 
values. One then checks the convergence of the resulting 
parameters and of the corresponding errors as the number of terms is
increased. After convergence, the
resulting errors include both statistical effects 
and systematic effects due
to truncation~\cite{LCDHMMT}.

To set the constraints, we shift and rescale
$F$ as follows, $\tilde F(x) = (F(x) - F_{\rm RMT})/(F_{\rm
  Poisson}-F_{\rm RMT})$, so that $\tilde F$ interpolates between 1
(localised/Poisson region) and 0 (delocalised/RMT region). The data
indicate that $\tilde F$ changes rapidly, monotonically and almost
linearly between 1 and 0 over a range $\delta x$. Any reasonable
definition of $\delta x$ has then to satisfy $1+\tilde F_1 \delta x \simeq
0$. Moreover, $\delta x$ provides a reasonable estimate 
of the radius of convergence $\rho$ of the series.
Furthermore, it is known that $(\tilde F_{n+1}/\tilde F_n) \rho \to 1$ as
$n\to\infty$, and so we expect $\tilde F_n \rho^n  \sim 1$ (at least
for large $n$). One then finds that $\tilde F_n/(-\tilde F_1)^n $  is
expected to be of order 1. This constraint was imposed rather loosely,
by asking $\tilde F_n/(-\tilde F_1)^n$ to be distributed according to
a Gaussian of zero mean and width $\sigma = 10$ for $n\ge 4$. We did
not impose any constraint on the coefficients $F_n$ with $n<4$, as
well as on $\nu$ and $\lambda_c$. The results of the constrained fits
converge rather rapidly as $n_{\rm max}$ is increased, see
Fig.~\ref{fig:binwidth_dep}. We went as far as $n_{\rm max}=9$, and we
used the corresponding results for the following analyses.  

The effects of the choice of bin size and fitting range
were checked by varying the bin size $\Delta\lambda$
and the width $w$ of the fitting range, which was centered
approximately at the critical point. The results show a slight
tendency of $\nu$ to decrease as $\Delta\lambda$ is decreased, 
but it is rather stable for $\Delta\lambda\cdot 10^{3}\lesssim 3$. 
There is also a slight tendency of $\nu$ to increase as $w$ is
decreased, becoming rather stable for $w\cdot 10^{2}\lesssim 3$. See
Fig.~\ref{fig:binwidth_dep}. To quote a single value for $\nu$, we
averaged the central values obtained for $1  \le \Delta\lambda\cdot
10^{3} \le  3$ and $2.6 \le w\cdot 10^{2} \le 3$. As the error is also
rather stable within these ranges, we quote its average as the final
error on $\nu$ for each choice of $L_{\rm min}$. We have checked that
other prescriptions (e.g., extrapolating to vanishing $w$ and/or
$\Delta\lambda$, or changing -- within reasonable bounds -- the ranges
of $w$ and $\Delta\lambda$ over which the final average is performed)
give consistent results within the errors. 

\begin{figure}[t]
  \centering
  \includegraphics[width=0.5\textwidth]{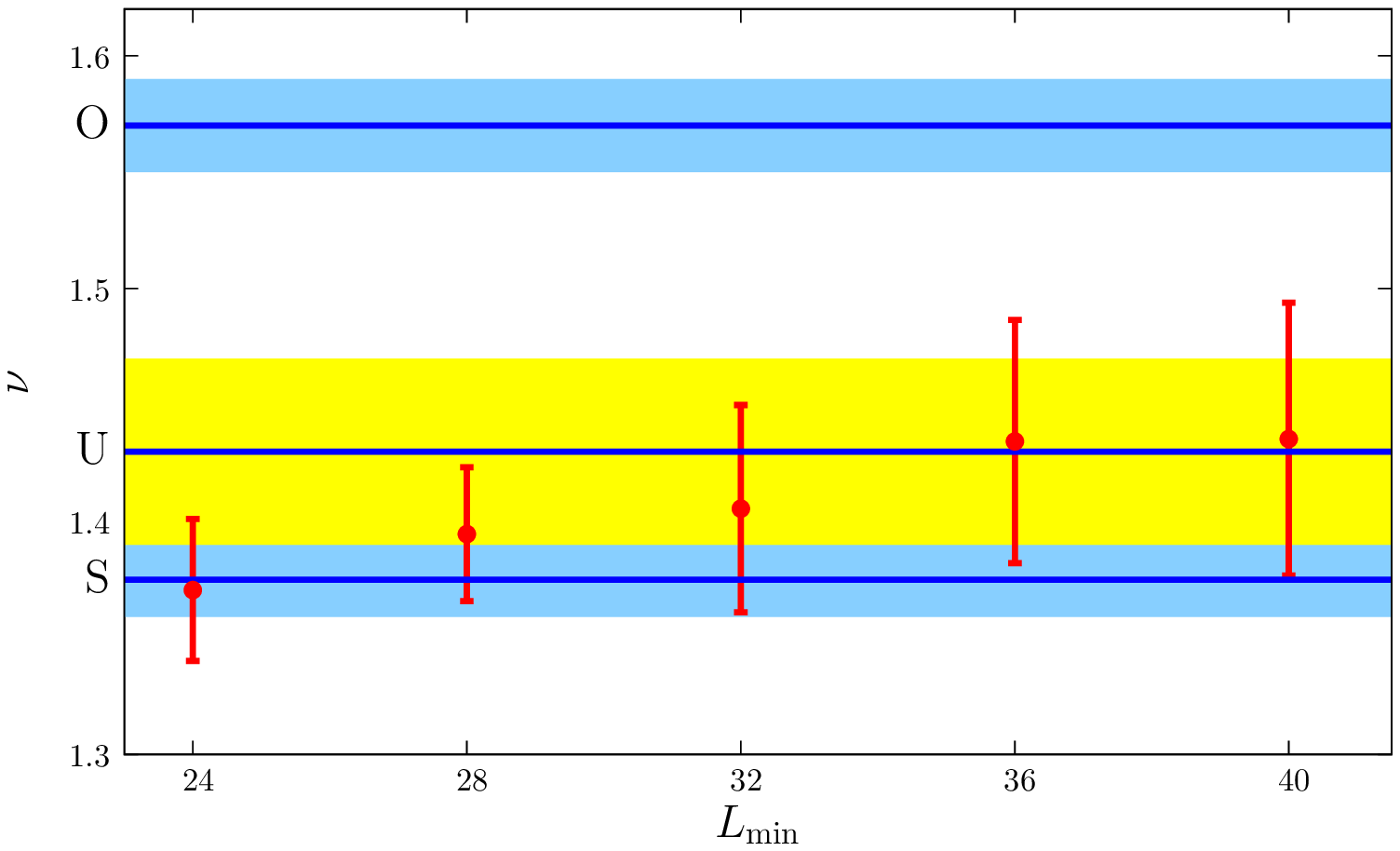}
  \hspace{\stretch{1}}
\includegraphics[width=0.48\textwidth]{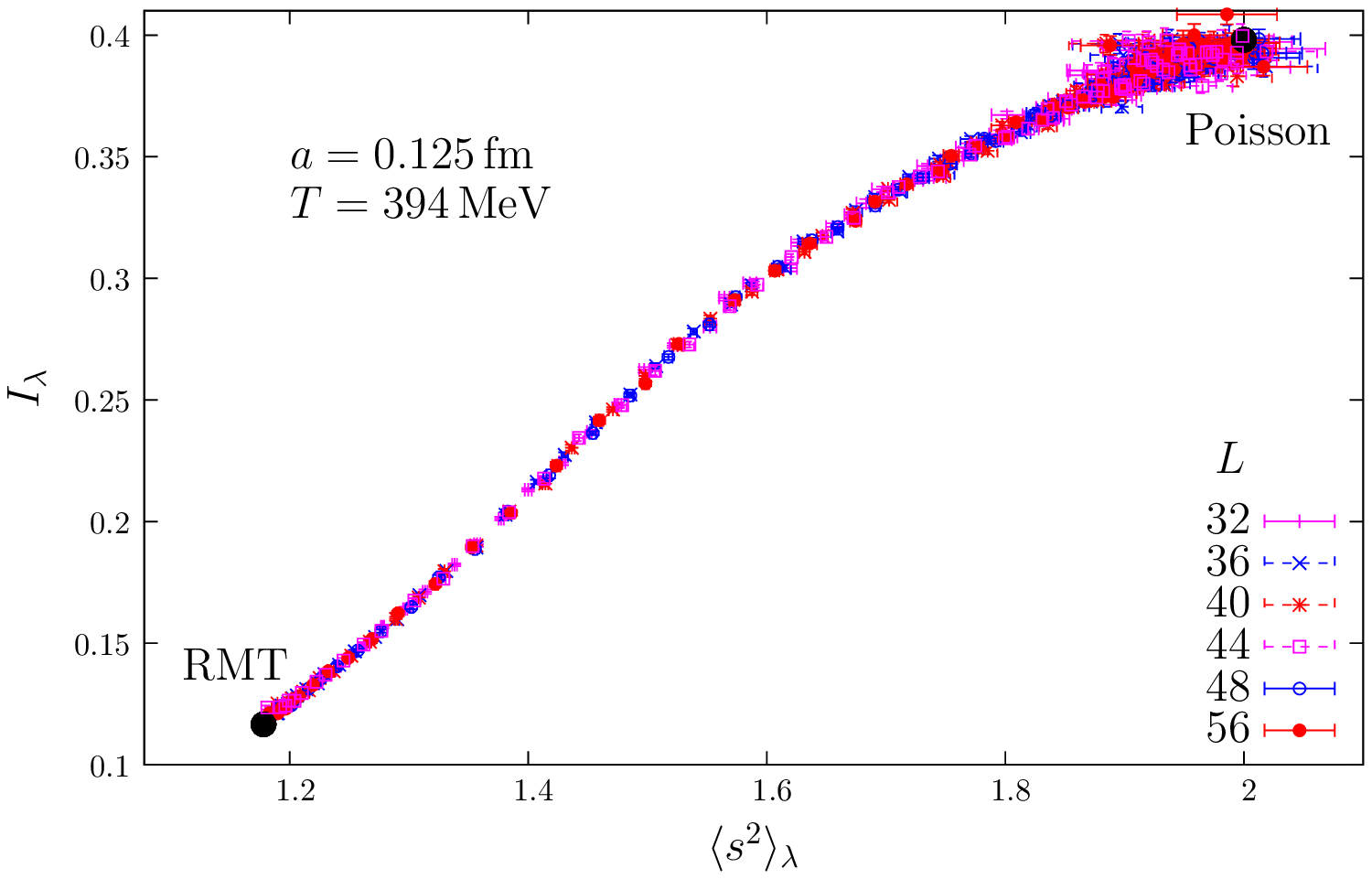}
  \caption{Dependence of the fitted value of $\nu$, averaged over
    $2.6\le w\cdot 10^2 \le 3.0$ and $1.0\le \Delta\lambda\cdot 10^3
    \le 3.0$, on $L_{\rm min}$. The values of $\nu$ obtained in
    the three symmetry classes of the 3D Anderson model (symplectic,
    $\nu_{\rm S}=1.375(16)$~\cite{nu_symp}, unitary $\nu_{\rm
      U}=1.43(4)$~\cite{nu_unitary} and orthogonal $\nu_{\rm
      O}=1.57(2)$~\cite{nu_orth}) 
    are shown for comparison together with their errors (left). Plot
    of $I_\lambda$ against $\la s^2\ra_\lambda$ for several lattice
    sizes (right).} 
  \label{fig:lmin_dep}
\end{figure}
Concerning finite size effects,
the fitted value of $\nu$ increases with $L_{\rm min}$,  stabilising
around $L_{\rm 
  min}=36$, see Fig.~\ref{fig:lmin_dep}.  This signals that our smallest 
volumes are still too small for one-parame\-ter scaling to work, and
that finite size corrections are still important there. On the other
hand, as the difference between the values obtained with $L_{\rm
  min}=36$ and $L_{\rm min}=40$ is much smaller than the statistical
error, one-parameter scaling works fine for our largest volumes. 

The value for the critical point $\lambda_c\simeq 0.336$ was obtained
through the same procedure described above. As a function of $L_{\rm
  min}$, the fitted value of $\lambda_c$ shows no systematic
dependence, and different choices of $L_{\rm min}$ give consistent
values 
within the errors. 

Our result for the critical exponent $\nu=1.43(6)$ is compatible with
the result obtained for the three-dimensional unitary Anderson model
$\nu_{\rm U}=1.43(4)$~\cite{nu_unitary}. This strongly suggests that
the transition 
found in the spectrum of the Dirac operator above $T_c$ is a true
Anderson-type phase transition, belonging to the same universality
class of the three-dimensional unitary Anderson model.

\section{Shape analysis}

From the point of view of random matrix models, Fig.~\ref{fig:2} shows
that the local spectral statistics along the spectrum are described by
one-parameter families of models, with spectral statistics
interpolating between Poisson and Wigner-Dyson along some path in the
space of probability distributions. To check if the appropriate
one-parameter family depends on the size of the lattice, 
one can simply plot a couple of parameters of the ULSD against each
other (thus projecting the path onto a two-dimensional plane in the
space of probability distributions): if points are seen to collapse on
a single curve, irrespectively of $L$, then the intermediate ULSDs lie
on a universal path in the space of probability
distributions~\cite{Varga}.  
  
In Fig.~\ref{fig:lmin_dep} we show $I_\lambda$ and $\la
s^2\ra_\lambda$ plotted against each other for several volumes, and we 
see that they indeed lie on a single curve. As $L$ is increased,
points corresponding to a given value of $\lambda$ flow
towards the Poisson or RMT ``fixed points'', while flowing away from
an  unstable fixed point corresponding to $\lambda_c$, where a 
different universality class for the spectral statistics is
expected. Similar plots made by changing $T$  and/or $a$ are
compatible with a similar universality of the path, but statistical
errors are still too large to reach a definitive conclusion.

The transition from Poisson to Wigner-Dyson behaviour in finite volume
is therefore expected to be described by a universal one-parameter
family of random matrix models~\cite{Nishigaki}. Comparing with
analogous results for the Anderson model, it turns out that the
spectral statistics at the critical point in the two models are
compatible~\cite{Nishigaki}.

\end{document}